\documentclass{article}
\usepackage{spconf,amsmath,graphicx}
\usepackage{multirow}
\usepackage{times}
\usepackage{enumitem}
\usepackage{latexsym}

\title{PDAF: A Phonetic Debiasing Attention Framework for Speaker Verification}
\name{\begin{tabular}[c]{@{}c@{}c@{}c@{}c@{}c@{}}
Massa Baali, Abdulhamid Aldoobi, Hira Dhamyal, Rita Singh, Bhiksha Raj 
\end{tabular}
}
\address{\begin{tabular}[c]{@{}c@{}c@{}c@{}c@{}c@{}}
 Carnegie Mellon University
\end{tabular}}

\begin{document}
\ninept
\maketitle
\begin{abstract}
Speaker verification systems are crucial for authenticating identity through voice. Traditionally, these systems focus on comparing feature vectors, overlooking the speech's content. However, this paper challenges this by highlighting the importance of phonetic dominance, a measure of the frequency or duration of phonemes, as a crucial cue in speaker verification. A novel Phoneme-Debiasing Attention Framework (PDAF) is introduced, integrating with existing attention frameworks to mitigate biases caused by phonetic dominance. PDAF adjusts the weighting for each phoneme and influences feature extraction, allowing for a more nuanced analysis of speech. This approach paves the way for more accurate and reliable identity authentication through voice. Furthermore, by employing various weighting strategies, we evaluate the influence of phonetic features on the efficacy of the speaker verification system.
\end{abstract}
\begin{keywords}
Phonetic, Speaker Verification, Speaker Identification, Phoneme Importance, Phonetic Debiasing 
\end{keywords}
\section{Introduction}
\label{sec:intro}
``Speaker verification'' is the problem of verifying if a voice recording belongs to a claimed identity \cite{campbell1997speaker}. This is typically done by building a model for the voice of the target speaker (the claimed identity), and evaluating the test recording against it \cite{reynolds1995automatic}. In the most common setting, the model itself just comprises one or more recordings from the target speaker. The test recording is compared to recordings from this ``gallery''. If the two recordings are ``close'' enough, i.e. if the similarity between them (as computed by some formula) exceeds a threshold, a match is declared and the test recording is verified as belonging to the speaker, if not it is rejected \cite{bai2020cosine}.

Speaker verification systems are known to be most accurate when they are \textit{text dependent}, i.e. when the test speaker is required to say a specific phrase, such as a passphrase, in order to be verified \cite{sarkar2016text}. Precise knowledge of what the speaker is supposed to say greatly aids in the accuracy of the verification.

On the other hand, arguably the far more common approach is \textit{text independent}, where the recordings are unrestricted in terms of what is spoken \cite{sarkar2016text}. The standard approach here is to derive a feature vector, or ``embedding'' from both the test and gallery recordings and compare them using a similarity metric such as the cosine similarity \cite{george2018analysis} or through statistical hypothesis testing under a framework such as PLDA \cite{wang2023incorporating}. Given the text-independent setting, the feature vectors are intentionally derived to be agnostic of the spoken content of the recording, focusing on the underlying speaker characteristics instead \cite{kinnunen2010overview}. Indeed, the feature extraction algorithms often include models that are explicitly trained to ignore the content of the recordings.

\begin{figure}
    \centering
    \includegraphics[width=0.5\linewidth]{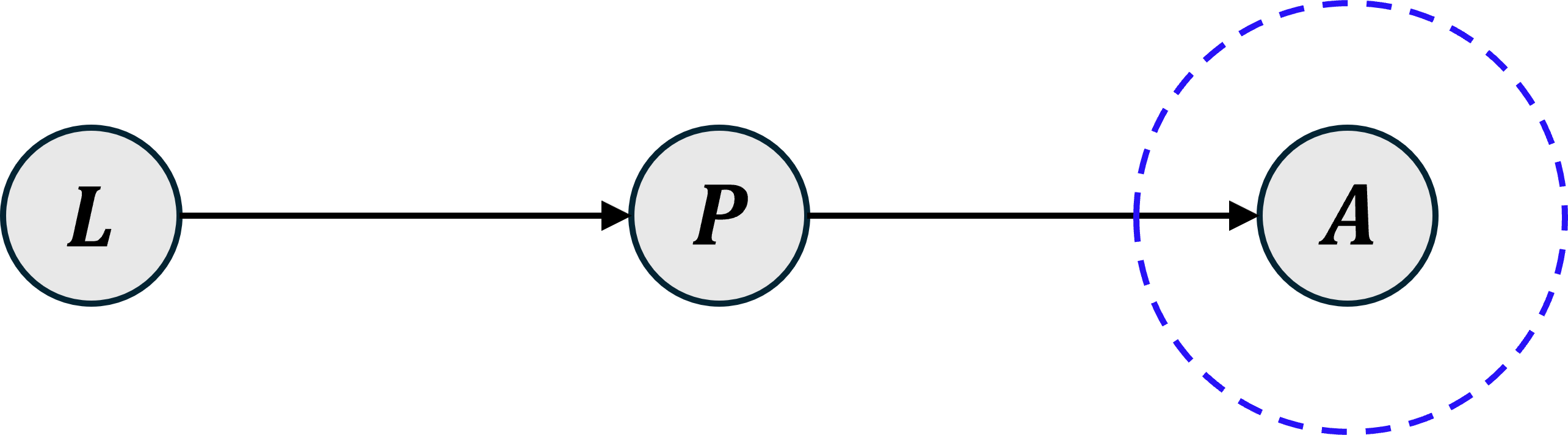}
    \caption{Statistical dependencies in the generation of a meaningful speech signal.  The lexical content $L$ determines the phonetic structure $P$, which in turn determines the acoustics $A$. Thus, any production of a signal actually represents the draws of all three variables. Content-agnostic verification systems, however, only consider the acoustics, which are the final variable $A$ (in the dotted circle), ignoring its conditioning on upstream variables $L$ and $P$.}
    \label{fig:dependency}
\end{figure}

This, however, ignores a key lesson learned from the text-dependent verification systems -- knowledge of what is spoken does indeed greatly enhance verification. Speech is in fact a highly structured signal, with higher-level semantic, lexical and phonetic structure, intended primarily to communicate information; the fact that it also carries information about the speaker is almost incidental to this primary purpose. The lexical and phonetic content govern the acoustic presentation, as illustrated in Figure \ref{fig:dependency}.  The production of a speech signal actually involves the joint production of the lexical, phonetic, and acoustic signals. Taking a statistical view of the process, each production is a draw from the joint distribution of lexical ($L$), phonetic ($P$), and acoustic ($A$) signals, or more simply, assuming an onto relationship between the phoneme sequence $P$ and the lexical content $L$ of the signal, from the joint distribution $\mathcal{P}(P,A)$ of phonetic and acoustic signals. The speech recording $A$ itself is just a partial view of this joint production. Content-agnostic models for feature extraction that consider only the acoustic signal effectively attempt to marginalize out the phonetic content, considering only the distribution of the acoustics, $\mathcal{P}(A) = E_P[\mathcal{P}(A|P)]$.

This form of marginalization is perfectly reasonable when training the model, when large amounts of varied training data are available enabling the marginalization. \textit{Inference}, however, is only performed on \textit{individual} recordings. An individual recording, as mentioned earlier, is a draw from $\mathcal{P}(P,A) = \mathcal{P}(P)\mathcal{P}(A|P)$. It effect, it comprises the draw of a phoneme sequence $\mathbf{P} \sim \mathcal{P}(P)$, followed by a draw of the acoustics from the conditional distribution, $\mathbf{A} \sim \mathcal{P}(A|\mathbf{P})$. Thus,  the specific phoneme sequence $\mathbf{P}$ acts as a conditioning prior on the acoustics, effectively biasing it. 

In this paper, we contend that the performance of text-independent speaker verification systems could be improved if the bias introduced by the phoneme sequence underlying any recording could be mitigated. 

Ideally, when sampling audio for the purpose of speaker verification, every phoneme in the language must be represented in proportion to its importance (for the purpose of speaker verification). However, by the nature of the problem (where the speaker's speech is unrestricted), we do not have control over the specific sequence of phonemes in the recording (although we may assume that the sequence itself is known, e.g. by manual transcription, or through automatic recognition), and thus cannot control the proportion of the recording representing any phoneme. Moreover, we also do not have an indication of the actual importance of the phonemes.

To tackle these problems, we propose a feature extraction framework that naturally \textit{debiases} phonemes through an approach inspired by importance sampling.  The contribution of each phoneme is rescaled by a factor that normalizes out (an estimate of) its actual occurrence probability, and scales it up to its target importance. Since the importance of each phoneme is unknown \textit{a priori}, we assume this to be uniform, although the framework also permits us to attempt to \textit{learn} their importance from data.  

In order to implement such a debiasing framework, the feature extraction itself must enable the consistent imposition of debiasing factors.  We use a self-attentive transformer-based model that operates on spectrographic representations of the speech signal. Within the model, the debiasing factors may be naturally applied in the form of attention masks that ensure that any reference to a segment of audio from a particular phoneme carries with it the corresponding debiasing weight. We call this model the ``Phoneme-Debiasing Attention Framework'', or PDAF.

Within this framework, we investigate a number of different ways of estimating the occurrence probability of phonemes to use within the debiasing factors. We also investigate whether the importance weights of the phonemes can be automatically estimated. We determine that by the appropriate choice of occurrence probability estimator and assuming uniform importance of phonemes, significant improvements in speaker verification performance can be obtained.

As an added benefit, the PDAF framework also enables the determination of the contribution of individual phonemes to speaker identity, through a process of ablation, whereby individual phonemes may be masked and the change in performance evaluated.  
While similar approaches have been attempted to derive phoneme contribution information from deep neural models earlier \cite{thebaud2024phonetic,liu2019introducing}, our approach is arguably cleaner, as we explain in Section \ref{sec:method}.
We find, through experiments across several models, that although individual phonemes do contribute to verification performance, the combined contribution is greater than the sum of the parts, suggesting that beyond just the acoustic signatures of individual phonemes, it is the manner in which they are co-articulated that may be more informative about the speaker. This opens up potential areas of investigation in the future.

The rest of the paper is organized as follows. In Section \ref{sec:prior}, we discuss prior work in areas related to our work. In Section \ref{sec:method}, we describe our Phoneme-Debiasing Attention Framework. In Section \ref{sec:expt}, we discuss our experimental evaluations. In Section \ref{sec:results}, we present our results and conclude with a discussion in Section \ref{sec:obs}.

\section{Background and related work}\label{sec:prior}
Speaker verification is a one-class classification problem (does the recording belong to the target identity or not?). It has traditionally been treated as \textit{hypothesis testing} -- identifying the confidence with which the null hypothesis $H_0$ that the test recording does \textit{not} belong to the target speaker can be rejected in favor of the alternate hypothesis $H_1$ that it does \cite{vera2015speaker}. This requires extraction of a sufficiently discriminative feature representation of the audio and explicit characterization of the probability distributions of the feature under both the null and alternate hypotheses \cite{hansen2015speaker}. The alternate and currently more successful approach treats this as an explicit one-class classification problem, where class membership is evaluated in terms of the similarity between the features derived from the test recording and a representative vector drawn from the class. The success of this approach depends critically on the representative nature of the feature vectors, or \textit{embeddings}, derived from the recordings. 

A large number of approaches have been proposed in the literature for the derivation of such embeddings, including Gaussian-mixture supervectors \cite{campbell2006support}, factor-analyzed supervectors \cite{dehak2010front}, features derived from speech recognizers \cite{ganapathy2009modulation}, and more recently neural-network based models including X-vectors \cite{snyder2018x}, D-vectors \cite{doddipatla2017speaker}, and transformer and conformer-architecture based models \cite{desplanques2020ecapa, gulati2020conformer, zhang2022mfa}. 

Most of these approaches, however, simply treat the speech recording either as a simple bag of vectors \cite{campbell2006support}, or as a generic time series \cite{desplanques2020ecapa, snyder2018x, doddipatla2017speaker}, without taking into consideration the lexical and phonetic conditioning factors behind the signal, as mentioned in Section \ref{sec:intro}.

Our work is based on the hypothesis that ignoring the underlying conditioning phonetic sequence could introduce biases and that explicitly accounting for them in the embedding computation could improve performance. We use a self-attention based framework to effect to reduce phonetic biases. In the process, we also obtain a framework that enables us to explicitly evaluate the speaker-discriminative characteristics of individual phonemes. Before proceeding, we briefly relate prior work from the field on these problems and contrast our work to them.

\textbf{The use of phonetic information in Speaker Verification}: It has been long known that explicit consideration of the phonemes underlying the speech signal is beneficial for text-independent speaker verification \cite{lei2014novel, rahman2018employing, liu2018speaker}. Most approaches, however, simply use the phonemes themselves directly as features. Zhou et al. \cite{zhou2019cnn} use sequences of phoneme embeddings as a separate feature stream, processed by a Convolutional Neural Network (CNN), and followed by an attention-based pooling to combine phonetic information with the acoustic signal. 
Li et al. \cite{li2019phonetic} employ a phonetic-correspondence based weighted similarity measure, where the similarity between acoustic frames corresponding to the same phoneme in the test and gallery recordings are assigned higher weight than those from different phonemes. This approach emphasizes frames with similar phonetic content but doesn't directly address the underlying bias introduced by the phoneme sequence itself and focuses on local context.  
Zhang et al. \cite{zhangintroducing} incorporate an auxiliary self-supervised branch into their feature extractor to detect phoneme boundaries, which information is then used to guide the feature extraction.
Liu et al. \cite{liu2019introducing} use a secondary phoneme prediction branch in their network to generate phoneme labels, which are then provided as additional inputs to a portion of the network that extracts the final feature vector.
Chen and bao \cite{chen2021phoneme}  use phoneme-specific TDNNs to better match the embeddings derived from a recording to its phonetic content.

In all of these works, although the phonetic information in the input is used, it is primarily used as an additional feature stream. The fact that the phonetic sequence is, in fact, a conditioning variable whose distribution could vary from recording to recording is not considered or accounted for. On the other hand, our debiasing framework aims to explicitly account for these variations, leading to a more robust solution. We have not identified any other work in the literature that attempts to do so.

\textbf{Attentive pooling for feature extraction}: A key component of our proposed framework is the use of attentive pooling to compute features. The use of self-attending networks for feature extraction in speaker verification systems is, in fact, well established, and a number of enhancements too have been proposed over it. Models such as the Conformer \cite{gulati2020conformer}, the MFA Conformer \cite{zhang2022mfa}, and their derivatives build on transformer architectures that employ self-attention and pooling as part of their architectures. Most feature extractors derive one vector per frame of speech; these must be pooled to get an utterance-level embedding, and here too attentive pooling is often used. Safari et al. \cite{safari2020self} proposed a transformer-based feature extractor followed by learned attentive pooling to extract features. 
Mun et al. \cite{mun2023frequency} introduces a selective kernel attention mechanism that uses an attention mechanism to dynamically choose the optimal kernel size in a CNN to extract features from the voice signal.
Like \cite{safari2020self}, Okabe et al. \cite{okabe2018attentive} propose a learned pooling, with the modification that it computes both the weighted mean and weighted second moments of frame-level features to derive utterance embeddings.  

Note that none of these approaches actually consider phoneme information or phonetic biases when computing the pooling; however, they do provide the kind of framework that we too employ in our work.

\textbf{Phoneme Importance}: A number of early studies have established that different phonemes carry differing levels of information about speaker identity \cite{sarkar2016text,kinnunen2010overview,zhangintroducing}. The studies themselves generally analyze isolated recordings of phonemes, typically either recorded by themselves or spliced out of longer recordings   \cite{eatock1994quantitative}, and evaluated either through human perception studies or using automated algorithms. Gallardo et al. \cite{gallardo2015importance} also analyze the contribution of different frequency bands to the speaker-discriminative ability of different phonemes. While these studies do establish that phonemes vary in their speaker-discriminative characteristics, they were generally performed on tiny amounts of data, from a small number of subjects, and their conclusions on the relative importance of phonemes or phoneme classes are sometimes contradictory. 

More recently, phoneme-contribution studies have been performed by analyzing the performance of neural-network based systems trained on large corpora.  Thebaud et al. \cite{thebaud2024phonetic} investigate the contribution of individual phonemes by slicing and splicing the instances of each phoneme to evaluate verification performance.
Kashyap et al. \cite{kashyap1976speaker} perform statistical hypothesis testing based speaker verifications on aggregated phoneme segments for each phoneme to identify their individual discriminativeness. 
Moez et al \cite{moez2016phonetic} go the other way and \textit{elide} phoneme categories from the audio, replacing it with noise, to determine how much each phoneme category contributes to verification.
Li et al. \cite{li2024phonemes} compute salience maps at the output of a stack of convolutions to determine which phonemes contribute most to performance.

Most of these approaches depend on excising out individual phonemes for evaluation, which not only greatly reduces the amount of data being used from any recording to evaluate it, but also introduces confounding edge effects.  Others work off the outputs of convolutions.  Convolutions ``smear'' information across the input in a manner that makes it hard to disentangle the contribution of any single phoneme at any location of the input, making any conclusions derived from them questionable.
Our solution avoids these problems by using a masking approach as described in Section \ref{sec:method}.

\section{Phoneme-Debiasing Attention Framework}
\label{sec:method}

The premise underlying our work is that any individual voice recording is actually a joint draw of \textit{two} variables, the underlying phoneme sequence $\mathbf{P}$ and the acoustic signal $\mathbf{A}$ from the joint distributions, $\mathcal{P}(A, P)$. Viewed as a sampling process, the phoneme sequence $\mathbf{P}$ is sampled from the phoneme sequence distribution: $\mathbf{P} \sim \mathcal{P}(P)$,  and the acoustics are sampled from the conditional distribution: $\mathbf{A} \sim \mathcal{P}(A | \mathbf{P})$.

When a content-agnostic feature extractor $f(A;\theta)$ (with parameter $\theta$) is trained on large volumes of training data, the training effectively minimizes $E_{P,A}[loss(f(A;\theta))]$, effectively marginalizing out the phoneme-sequence variable $P$. This feature extractor is now optimized for the global distribution of phoneme sequences.

At inference time, however, we only operate on individual recordings, drawn from $\mathcal{P}(A|P)$. The challenge is that the \textit{a priori} distribution of phonemes $\mathcal{P}(P)$ may differ even from utterance to utterance. $f(A;\theta)$ is no longer optimal for the utterance. To give a concrete example, if the phoneme /AH/ dominates in the training data, but the test utterance comprises primarily the phoneme /IY/, $f(A;\theta)$ optimized on the training set cannot be expected to be optimal for the test utterance. The phoneme priors of the test utterance impose a bias that is not applicable to the model. We may expect to gain by debiasing it.

Our solution is to modify the feature extraction model to $E_P[f(A|P;\theta)]$, where $P \sim \mathcal{P}(P)$, the global (training) distribution of phonemes. The above formula seems to imply phoneme-specific feature extraction, a complication we would like to avoid. Instead, we restate the above as 
\begin{equation}   
f(A|P;\theta) = \left[ E_{P_{t'}}[f_t(A_{t'}; \theta)]; t = 1\cdots T\right]
\end{equation}
where $A = A_1, A_2, \cdots, A_T$ and $P = P_1, P_2,\cdots, P_T$ are assumed to be sequences of $T$ vectors,  $A_t$ is the $t^{\rm th}$ vector in the sequence $A$, and $P_t$ is its underlying phoneme, and $f_t()$ is a time-specific function applied to $A_t$. Note that although $f(A_t;\theta)$ does not explicitly denote the phoneme, it is still nonetheless implicitly dependent on the phoneme sequence $P$ since each $A_t$ depends directly on its own phoneme $P_t$, as well as implicitly on the rest of the phoneme sequence through context effects. 

The above equation assumes that a single local feature vector is computed at each time $t$, which is itself a phoneme-dependent expectation over itself and its siblings. When the phoneme sequence $P$ is drawn from the global training distribution $\mathcal{P}(P)$, the expectation $E_{P_t}[.]$ simply becomes an averaging operation:  
\[
f(A|P;\theta) = \left[ \frac{1}{T} \sum_{t'} f_{t}(A_{t'};\theta) \right]
\].

However, when the phoneme sequence $P$ underlying $A$ comes from a different distribution $\hat{\mathcal{P}}(P)$, we must make adjustments to extract the correct feature \cite{gelman1995bayesian}:
\begin{equation}
f(A|P;\theta) = \left[\sum_{t'} \frac{\mathcal{P}(P_{t'})}{\hat{\mathcal{P}}(P_{t'})} f_t(A_{t'};\theta); t = 1\cdots T\right]
\label{eq:debias}
\end{equation}
where, as before, $\mathcal{P}(P_t)$ is the global probability of $P_t$, and $\hat{\mathcal{P}}(P_t)$ is the \textit{local} probability distribution of phonemes with the utterance.  Note that in the above equation, the phonemes $P_t$ in the sequence $P$ are assumed IID as a simplification; however, this is not essential. More detailed time-series distributions too can be accommodated within the expectation.

In our framework, we have modelled $f_t()$ using a self-attention framework \cite{vaswani2017attention}:
\[
f_t(A_{t'};\theta) = \sum_{t'} w(t,t')g(A_{t'})
\]
where $w(t,t')$ are attention weights, and $g(A_t)$ is a strictly local transformation applied to $A_t$.  When combined with Equation \ref{eq:debias}, we get
\begin{equation}
    f(A|P;\theta) =\left[\sum_{t'} \frac{\mathcal{P}(P_{t'})}{\hat{\mathcal{P}}(P_{t'})} w(t,t')g(A_{t'};\theta); t = 1\cdots T \right]
\label{eq:finaldebias}
\end{equation}

The above framework is very easily implemented within a self-attending transformer block. Our actual model comprises several such blocks. The two questions that remain to be answered are (a) What is the global distribution $\mathcal{P}(P)$, and (b) what is the local distribution $\hat{\mathcal{P}}(P)$. Note that the global distribution $\mathcal{P}(P)$ does not necessarily refer to the actual global distribution of the phonemes, but could be any distribution capturing the relative importance of phonemes. Our primary model assumes that  $\mathcal{P}(P)$ is uniform; however, we also consider the possibility of learning it. Note that since the ``global'' phoneme probability distribution is unlikely to match any of the recordings in the training data, Equation \ref{eq:finaldebias} is used to process training data utterances as well.

Below, we discuss our actual implementation of the model as a transformer block, and various ways of estimating $\hat{\mathcal{P}}(P)$.

\subsection{Implementing the model}

For each utterance we extract the phonemes and their durations from the speech and transcriptions using an unsupervised phonetic aligner \cite{zhu2022phone}. These phonemes form the basis for calculating probabilities. The model itself is implemented as a stack of multi-head self-attention blocks operating on the Mel spectrogram of the speech.  Within each block, assuming $\mathcal{P}(P)$ to be uniform, we introduce a phonetic scaled dot product that adjusts the attention scores by subtracting the logarithm of the estimated phoneme probabilities $\hat{\mathcal{P}}(P)$. Additionally, the phoneme probabilities are aligned with an attention mask wherein we specifically mask the silence index. Finally, after processing the input with the multi-head self-attention blocks, we use an attentive pooling mechanism \cite{safari2020self} to compute the mean and standard deviation of the encoded features, which are concatenated to create an utterance-level embedding. These features are then fed through a batch normalization layer followed by a ReLU activation, transforming them into an overall speaker representation. 

To train the network, this speaker representation is subsequently fed into a single classification layer with a softmax that outputs the speaker posteriors. The model is trained to minimize classification loss over the training set speakers.  During verification, similarity scores are computed between embeddings using a cosine similarity.

\subsubsection{Phonetic Aligner}
We calculate the phoneme logits using an unsupervised phonetic aligner with a vocabulary of 40 phonemes and one index for silence. We kept the silence index even though it's masked during self-attention same as the padded parts. Although we already had the actual transcriptions, the unsupervised phonetic aligner could still produce the phoneme logits without the ground-truth transcriptions.
\subsubsection{Multi-Head Self-Attention}
The actual PDAF model is implemented in the form of multi-head self-attention blocks. The core of self-attention involves calculating attention scores that include phonetic priors. This is done through a scaled dot product with phonetic attention masking.
%
Our approach directly incorporates phoneme probabilities ($\hat{\mathcal{P}}$) into the attention scores as below. Here $\hat{\mathcal{P}}$ is a vector representing the estimate of the actual probability distribution of phonemes in the recording.  
\begin{equation}
    Attention(q_i, k_j) = softmax\left( \frac{Q_i K_j^T} {\sqrt{d_k}} - \lambda\log({\hat{\mathcal{P}}}_j)\right) 
    \label{eq:scaled}
\end{equation}
Subtracting the logarithm of the phoneme probability ($\log({\hat{\mathcal{P}}}_j)$) from the scaled dot-product attention scores effectively scales down the probability of each phoneme by its estimated local-occurrence probability, as required by Equation \ref{eq:finaldebias}.

\subsection{Estimating $\hat{\mathcal{P}}$} \label{sec:formulas}
We introduce different methods for computing the $\hat{\mathcal{P}}$ in the phonetic attention algorithm.
\begin{itemize}[leftmargin=*]
\item{Phoneme Overall Occurrence Probability (POP)}

This formula calculates the probability ($\hat{\mathcal{P}}$) of encountering a specific phoneme ($\phi$) within the entire dataset:

\begin{equation}
\hat{\mathcal{P}}_{\phi}^{overall} = \frac{N_{\phi}^{total}} {\sum_{\phi'} N_{\phi'}^ {total}}
\label{eq:POP}
\end{equation}
$N_\phi^{total}$: Total number of instances of phoneme $\phi$ across the entire dataset.

${\sum_{\phi'} N_{\phi'}^{total}}$: Summation over all possible phonemes ($\phi'$) of their total instance counts $(N_{\phi'}^{total}).$

\item{Phoneme Utterance Occurrence Probability (PUP)}

This formula calculates the probability ($\hat{\mathcal{P}}$) of encountering a specific phoneme ($\phi$) within a specific utterance (U):
\begin{equation}
\hat{\mathcal{P}}_{\phi}^{U} = \frac{N_{\phi}^{U}} {\sum_{\phi'} N_{\phi'}^U}
\label{eq:PUP}
\end{equation}

$N_\phi^{U}$: Number of instances of phoneme $\phi$ within the specific utterance U.

${\sum_{\phi'} N_{\phi'}^U}$: Summation over all possible phonemes ($\phi'$) within the utterance U of their instance counts $(N_{\phi'}^U).$

\item{Phoneme Frame Probability (PFP)}

This formula calculates the probability of frames (F) associated with a specific phoneme ($\phi$) within the entire dataset:
\begin{equation}    
\hat{\mathcal{P}}_{\phi}^{frame} = \frac{N_{\phi}^{frame,total}} {\sum_{\phi'} N_{\phi'}^ {frame,total}}
\label{eq:PFP}
\end{equation}
$N_{\phi}^{frame,total}$: Total number of frames associated with phoneme $\phi$ across the entire dataset.

$\sum_{\phi'} N_{\phi'}^ {frame,total}$: Summation over all possible phonemes ($\phi'$) of their total frame counts $(N_{\phi'}^ {frame,total})$

\item{Phoneme Frame Utterance Probability (FUP)}
This formula calculates the probability ($\hat{\mathcal{P}}$) of a specific frame (F) within an utterance (U) belonging to a specific phoneme ($\phi$):

\begin{equation}
\hat{\mathcal{P}}_{\phi}^{F,U} = \frac{N_{\phi}^{frame,U}} {\sum_{\phi'} N_{\phi'}^
{frame,U}}
\label{eq:FUP} 
\end{equation}

$N_{\phi}^{frame,U}$: Number of frames associated with phoneme $\phi$ within the specific utterance U.

${\sum_{\phi'} N_{\phi'}^{frame,U}}$: Summation over all possible phonemes ($\phi'$) within the utterance U of their frame counts $(N_{\phi'}^{frame,U}).$



\item{Learned Phoneme Weights}: In the previously described variants the global phoneme prior $\mathcal{P}$ was assumed to be uniform.  Alternately they can be directly learned as a phoneme-specific set of weights, during the training process. In this case, explicit modeling of $\hat{\mathcal{P}}$ is not required.
\end{itemize}

\subsubsection{Attention Masking}
We also incorporate an attention mask (M) that specifically masks the silence index in the spectrogram. This masking speech representation is aligned with the phonemes. This ensures the model doesn't attend to irrelevant padded regions or silence segments. 
\subsubsection{Masking Individual Phonemes}\label{sssec:mask}
During Inference, individual phonemes can be masked through an additional mask that zeroes out only the segments of the input corresponding to the phoneme. This is used in our experiments to determine phoneme importance.

\subsection{Self-Attentive Pooling with Mean and STD
}
The multi-head self-attention layers result in a sequence of feature vectors. These are combined into a single vector comprising the weighted component-wise mean and standard deviation of the vectors, computed using the self-attentive pooling method described in \cite{safari2020self}. 

\subsection{Speaker Embeddings}
The concatenated features from the pooling layer are transformed into an overall speaker representation by feeding them through a batch normalization layer followed by a ReLU activation function. A single classification layer comprising a linear layer followed by a softmax then takes this speaker representation as input. The softmax layer outputs the speaker posteriors, which represent the probability distribution of the speaker identities in the training set. The model is trained to minimize classification error on the training-set speakers.

Once trained, the output of the RELU activation comprises the speaker embeddings.
\begin{figure}[htb]
	
	\begin{minipage}[b]{1.0\linewidth}
		\centering
		\centerline{\includegraphics[width=8.5cm]{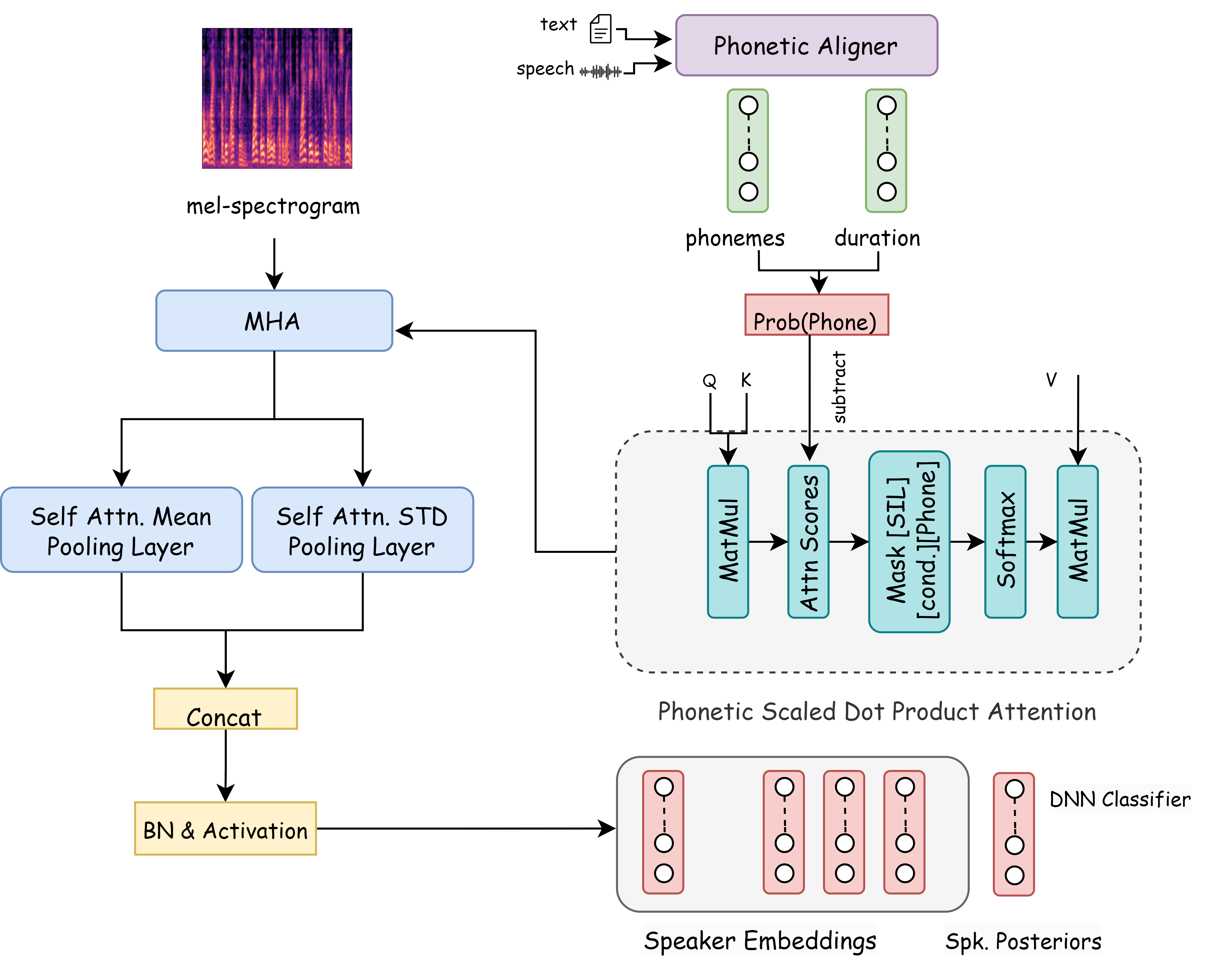}}
	\end{minipage}
	\caption{PDAF proposed phonetic integration for speaker recognition.}
	\label{fig:res}
\end{figure}
	
\begin{figure}[htb]
	
	\begin{minipage}[b]{1.0\linewidth}
		\centering
		\centerline{\includegraphics[width=8.5cm]{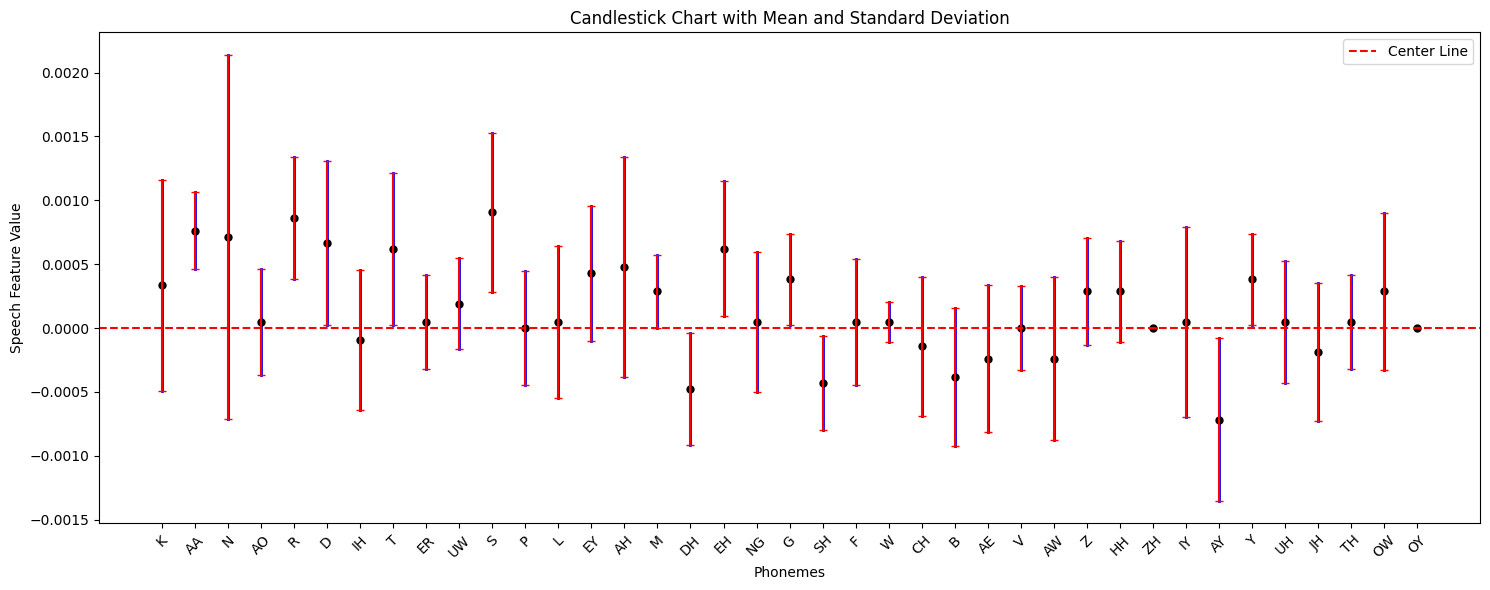}}
	\end{minipage}
	\caption{Mean and STD of the change in EER when a phoneme is masked, computed over 6 models.}
	\label{fig:candle}
\end{figure}

\section{Experiments}
\label{sec:expt}
\subsection{Datasets and Preprocessing}
\label{sec:data}
We conducted our experiments using 100h LibriSpeech. 28,539 utterances for the training and 2,703 utterances for the validation. Initially, we transform the input audio waveform into a sequence of 128-dimensional log Mel filterbank (fbank) features. These features are computed utilizing a 25ms Hamming window with a stride of 10ms, culminating in a spectrogram that serves as the input for the multihead self-attention mechanism.

We employ a phonetic aligner to extract the phonemes and their corresponding durations in each audio sample. Subsequently, we extract: attention masking vectors, POP vectors, and PFP vectors. These vectors are derived using the formulae detailed in Section \ref{sec:formulas}.

\subsection{Experimental Setup}
\label{sec:exp}
Figure~\ref{fig:res} illustrates the main components of our framework. This section details the experiments we conducted to evaluate different phoneme probability integration methods.

We trained four models with different local phoneme probability integration strategies:

\textbf{Baseline (no-weighting)}: We train a model that serves as a reference point, not incorporating any phoneme probabilities ($\lambda$ = 0). It takes the attention speech mask and spectrogram as input.

\textbf{POP}: This model utilizes pre-computed probabilities reflecting the overall phoneme distribution across the entire dataset. The input includes POP vectors, the attention speech mask, and the aligned spectrogram.

\textbf{PFP}: Similar to (2), this model uses pre-computed probabilities, but here it considers the probability of frames belonging to each phoneme in the entire dataset. The input includes PFP vectors, the attention speech mask, and the aligned spectrogram.

\textbf{Learnable Weighted Phoneme}: Unlike the previous models, this one doesn't pre-compute probabilities. Instead, it learns phoneme weights directly within the self-attention mechanism. The input includes a random initial learnable weights vector, the attention speech mask, and the aligned spectrogram.

\textbf{Model Configuration}: For all experiments, speech samples and speaker IDs were fed into the multi-head self-attention layer. We have configured the key and value-vector dimensions to be 32, with a speaker embedding dimension set at 1024, and the number of attention heads is set to 8. The multi-head attention components within the self-attention layers utilize these 8 attention heads with an attention dimension of 128. For the feedforward module of the self-attention layers, as well as the proposed feedforward encoder layers, we have set the corresponding dimension to 1024. A batch normalization layer is followed by a ReLU activation function. 
All models are implemented using PyTorch \cite{paszke2019pytorch}, leveraging the Adam optimizer with an initial learning rate of 0.001. This learning rate is halved every 4 epochs, in line with previous work \cite{zhang2022mfa}. To mitigate overfitting, we have set the weight decay to 1e-7 and employed a linear warmup for the first 2000 steps. The batch size is maintained at 100, and trained for 30 epochs. All experiments were conducted using NVIDIA A100 GPUs.
\section{Results and Analysis}
\label{sec:results}
Our study evaluates the significance of phonemes in speaker identification using the LibriSpeech test set. We initiate this process by generating pairs for speaker verification and selecting speech samples from various speakers. Each pair is composed of two segments: one 'same-speaker' segment, where both parts are from the same individual, and one 'different-speaker' segment, with parts from two distinct individuals. We utilize cosine distance for scoring purposes.
The systems' performances are computed using the Equal Error Rate (EER). Further, we extract speaker embeddings 
across different models to assess the speaker verification system's performance. During inference, for both the POP and PFP models, we explore the feasibility of calculating probabilities as indicated in the PUP (Equation \ref{eq:PUP}) and FUP (Equation \ref{eq:FUP}). Typically, in inference mode, we lack access to the entire dataset's statistics, making it preferable to compute phoneme and frame probabilities within individual utterances.
The results, presented in Table \ref{tab:eer_main}, reveal that the best results are obtained with the PUP model, which assumes uniform global priors $\mathcal{P}$,  employs the global phoneme-level estimate of phoneme probabilities as $\hat{\mathcal{P}}$ on the training data, and local (utterance level) phoneme probabilities for $\hat{\mathcal{P}}$ on test data. \textit{Estimating} phoneme weights proved to be marginally inferior to the pre-computed phoneme probabilities.
\begin{table}[htbp]
\caption{EER(\%) comparison among different methods}
\label{tab:eer_main}
\centering
\begin{tabular}{lcccp{1.5cm}}
\hline
\textbf{ID} & \textbf{Method} & \textbf{EER} \\ \hline
1 & Baseline (no-weighting) &  6.69 \\ 
2 & POP & 6.43 \\ 
3 & PFP & 7.06 \\ 
4 & PUP &  \textbf{6.29} \\  
5 & FUP &  6.98  \\ 
6 & Learnable Weighted Phone & 7.27 \\ \hline
\end{tabular}
\end{table}

Table \ref{tab:eer_align} demonstrates our approach's ability to handle speech samples without corresponding text. We use the unsupervised phonetic aligner, which utilizes a phonetic recognizer to predict phonemes directly from the audio. Interestingly, the table demonstrates that this textless alignment approach achieves slightly better performance compared to forced alignments, which require text input alongside the speech samples.
\begin{table}[htbp]
\caption{\footnotesize{Comparison of PUP EER for Different Alignment}}
\label{tab:eer_align}
\centering
\begin{tabular}{lcccp{1.5cm}}
\hline
\textbf{ID} & \textbf{Alignment} & \textbf{EER} \\ \hline
1 & Forced Alignment &  6.29 \\ 
2 & Textless Alignment & \textbf{6.20} \\ \hline
\end{tabular}
\end{table}

\begin{table}[htbp]
\caption{\footnotesize{Phonetic Classes from the ARPABET\cite{garofolo1993darpa}}}
\label{tab:phone_class}
\centering
{
\scalebox{0.80}{
\begin{tabular}{l|cccp{1.5cm}}
\hline
\textbf{Category} & \textbf{Symbol} \\ \hline
vowels & \vtop{\hbox{\strut AA, AE, AH, AO, AW, AY,}\hbox{\strut EH, ER, EY, IH, IY, OW, OY, UH, UW }} \\ 
\hline
\multicolumn{2}{l}{consonants}     \\ \hline
Fricative & \vtop{\hbox{\strut F, V, TH, DH}} \\
Stop & \vtop{\hbox{\strut P, B, T, D, K, G}} \\
Nasal & \vtop{\hbox{\strut M, N, NG}} \\
Sibilant & \vtop{\hbox{\strut S, Z, SH, ZH}} \\
Affricate & \vtop{\hbox{\strut CH, JH}} \\
Approximant & \vtop{\hbox{\strut W, R, Y}} \\
Lateral & \vtop{\hbox{\strut L}} 
\\
\hline
\end{tabular}
}}
\end{table}

\begin{table}[htbp]
\caption{EER(\%) comparison for PUP by masking different phoneme classes}
\label{tab:eer_mask_phnclass}
\centering
{
\scalebox{0.85}{
\begin{tabular}{lcccp{1.5cm}}
\hline
\textbf{Method} & \textbf{EER} \\ \hline
No Masking &  6.29 \\ 
Vowels & 7.18 \\ 
Fricative & 6.15 \\ 
Stop &  6.70  \\  
Nasal &  6.35  \\ 
Sibilant & 6.43 \\ 
Affricate &  6.20  \\ 
Approximant &  6.45  \\ 
Lateral & 6.20   \\ \hline
\end{tabular}
}}
\end{table}

Finally, to gain a deeper understanding of how individual phonemes influence speaker verification performance, for each of our models we mask out individual phonemes as described in Section \ref{sssec:mask}, and compute the EER. Figure~\ref{fig:candle} presents the change in EER from masking phonemes, using candlestick charts. Arguably, this approach is cleaner for evaluating phoneme contributions, than other methods that either slice out phonemes or attempt to identify importance after convolutive mixing of phonemes. Yet, besides a small number of phonemes for which removing the phoneme clearly hurts performance, the results are inconclusive for others. 

Assuming phoneme sparsity within individual recordings, we subsequently employ the same method over classes of phonemes.
We follow the ARPABET\cite{garofolo1993darpa} to determine the class of each phoneme. This involved masking each phoneme class (along with silence) within the attention mask for each model separately in the inference. The resulting EERs reflect the impact of removing a specific phoneme class on speaker verification performance. Table \ref{tab:eer_mask_phnclass} shows that the vowels class has the highest impact on the performance of the speaker verification system followed by the stop class and aproximant class.    

\section{Observations and conclusions}
\label{sec:obs}
The results clearly indicate that debiasing phoneme probabilities results in a clear improvement in performance, with a gain of up to 6\% relative for PUP. In particular, the specific model that provides the greatest improvement uses global training statistics to normalize training data, and local phoneme statistics to normalize test utterances, validating our original hypothesis that distribution shifts between test utterances and training data could affect performance.  From Table \ref{tab:eer_align} we note that we do not need \textit{a priori} knowledge of the phoneme sequences in the data for such gains; they are retained even when the sequence is estimated. Our approach does not exclude the ability to incorporate other phoneme-dependent processing schemes from the literature, indicating that further gains may be made by combining them.

Experiments on the importance of phonemes seem to support studies such as \cite{kinkiri2020phonemes, singh2016formant,singh2017voice} on the surprising importance of stops and sibilants on speaker verification, although on other phonemes and phoneme classes the results remain inconclusive. One must conclude that it is not merely the individual phonemes, but the \textit{co-articulation} between phonemes, which has not been evaluated here, that carry speaker information.  Fortunately, the proposed PDAF framework enables us to run studies on much larger corpora than ever before attempted, at minimal compute overhead, to obtain greater insight into the issue.  This remains future work.


\bibliographystyle{IEEEbib}
\bibliography{strings,refs}

\end{document}